\documentclass[aps,superscriptaddress,notitlepage,nofootinbib,twocolumn]{revtex4-2}
\usepackage[T1]{fontenc}
\usepackage[utf8]{inputenc}
\usepackage[english]{babel}
\usepackage{graphicx,microtype,amsfonts,amssymb,amstext,amsmath,mathtools,cancel,physics,accents,scalerel,xcolor}
\usepackage[hidelinks]{hyperref}
\newcommand{\stg}[1]{\accentset{\scaleto{\diamond}{3.5pt}}{#1}}
\DeclareMathOperator{\diag}{diag}

\begin{document}
\title{The Pre-geometric Origin of Geometric Trinity of Gravity}

\author{Salvatore Capozziello}
    \email{capozziello@na.infn.it}
    \affiliation{Dipartimento di Fisica ``E.\ Pancini'', Università di Napoli ``Federico II'', Via Cinthia 9, I-80126 Napoli, Italy}
    \affiliation{Scuola Superiore Meridionale, Via Mezzocannone 4, I-80134, Napoli, Italy}
    \affiliation{INFN Sezione di Napoli, Complesso Universitario di Monte Sant'Angelo, Edificio 6, Via Cinthia, I-80126, Napoli, Italy}

\author{Giuseppe Meluccio}
    \email{giuseppe.meluccio-ssm@unina.it}
    \affiliation{Scuola Superiore Meridionale, Via Mezzocannone 4, I-80134, Napoli, Italy}
    \affiliation{INFN Sezione di Napoli, Complesso Universitario di Monte Sant'Angelo, Edificio 6, Via Cinthia, I-80126, Napoli, Italy}
    \affiliation{Department of Physics, Loyola Marymount University, 1 LMU Drive, 90045, Los Angeles, California, USA}

\begin{abstract}
    The so-called Geometric Trinity of Gravity is based on three distinct geometric features of spacetime, i.e.\ curvature, torsion and non-metricity, which give rise to equivalent dynamics for General Relativity (GR), Teleparallel Equivalent of General Relativity (TEGR) and Symmetric Teleparallel Equivalent of General Relativity (STEGR). Pre-geometric gravity, on the other hand, offers a unifying framework from which all metric-affine theories can emerge. Starting from a gauge formulation \textit{à la} Yang--Mills with a Higgs-like field, a mechanism of spontaneous symmetry breaking can give rise to an effective metric as well as to the classical dynamics of the gravitational field. In particular, the emergence of gravity in the spontaneously broken phase is shown to be consistent with all the different formulations of the Geometric Trinity of Gravity, in terms both of actions and of gauge choices for the affine connection. This general result is achieved by deriving and analysing suitable expressions in the unbroken phase for pre-geometric actions and for pre-geometric gauge-fixing conditions respectively.
\end{abstract}

\maketitle

\section{Introduction and motivations}
Metric-affine theories of gravity consider extensions of the geometric framework adopted in GR: other than treating the affine connection as independent of the metric, in fact, they also allow for it to have more general properties than those of the Levi-Civita connection assumed in GR \cite{hehl:metric,hehl:solutions,hehl:space-time,hehl:relaxation,hehl:spin,obukhov:approach,vitagliano:dynamics,sotiriou:f(R),capozziello:f(R),capozziello:f(T),barrientos:metric,delhom-latorre:observable,iosifidis:metric,yang:integral,kibaroglu:maxwell,capozziello:affine}. The main motivations behind this area of research are pinpointing the true physical degrees of freedom of the gravitational field \cite{sauro} and elucidating its fundamental relation to the geometry of spacetime. Within this context, the Geometric Trinity of Gravity refers to a set of three distinct yet analogue theories of gravity: GR, TEGR and STEGR. The dynamical equivalence of these theories can be proven at different levels \cite{capozziello:comparing,capozziello:equivalence, battista}: the variational principle, the field equations and their solutions. Each of these formalisms describes gravity as the dynamical effect of some features of spacetime, which is seen as a differentiable manifold endowed with both metric and affine structures. However, the precise geometric source of gravity is different in each of these approaches:
\begin{enumerate}
    \item curvature in GR,
    \item torsion in TEGR,
    \item non-metricity in STEGR.
\end{enumerate}
Curvature is a measure of how parallel transport around a loop fails to return a vector to its original direction. Torsion is instead a measure of how parallel transport around an infinitesimal parallelogram fails to close, thus manifesting as a sort of `twisting' of spacetime. As for non-metricity, it is a measure of how parallel transport alters the length of vectors. Given that curvature, torsion and non-metricity have very different definitions and meanings in Riemannian geometry, it might then seem puzzling that it is even possible to construct a different but equivalent gravitational theory with each of them, i.e.\ the Geometric Trinity of Gravity \cite{Lavinia1}. The main goal of this work is to show that such differences do not stem from the specific \emph{geometric} rules determining the very fabric of spacetime, rather they are emergent properties of its fundamentally \emph{pre-geometric} nature.

Pre-geometric gravity is a proposed theoretical framework for recasting the gravitational interaction in a form akin to that of Yang--Mills theories, with a view to pursue an alternative path in the quest for Quantum Gravity \cite{macdowell:unified,stelle:de-sitter,pagels:gravitational,wilczek:gauge,wise:cartan,westman:introduction,tresguerres:dynamically,gallagher:pregeometric,palumbo:emergent,addazi:pre-geometry,addazi:hamiltonian,meluccio:pregeometric,addazi:emergence,addazi:topological,addazi:prospective,addazi:holographic,addazi:solution}. In their most basic form, pre-geometric theories of gravity are formulated on a four-dimensional differentiable manifold devoid of a metric structure, with the fundamental degrees of freedom being those of the gauge potential of the (anti-)de Sitter group. The dynamics of an additional Higgs-like field can then translate the initial formalism \textit{à la} Yang--Mills into that of the Einstein--Cartan theory of gravity, via a mechanism of spontaneous symmetry breaking (SSB). The symmetry-breaking pattern thus reduces the fundamental gauge symmetry of spacetime from $SO(1,4)$ or $SO(2,3)$ to $SO(1,3)$, at an energy scale which is expected to be Planckian. Not only, the complete emergence of the gravitational theory in the spontaneously broken phase is accomplished by recovering the Einstein--Hilbert action, the Einstein and Cartan field equations, the energy scales of gravity, the founding principles of GR and possibly also all the couplings to the matter fields of the Standard Model. Such a radical approach thereby deposes the metric from the central role it holds in Einstein gravity exactly at the energy scale where GR is expected to break down, leaving us to ponder the peculiarities of a primitive spacetime deprived of the concepts of distance, angle and causality.

In the models studied so far in literature, the correspondence between a pre-geometric, gauge theory and a geometric, gravitational one was established only for the Einstein--Hilbert action and the cosmological constant term (with one case recovering the Gauss--Bonnet term too \cite{macdowell:unified,addazi:pre-geometry}). Instead, little to no attention was devoted to analysing the characteristics of the emergent affine connection nor to studying the possibility of reconstructing more general gravitational actions. Therefore, our objective is twofold.
\begin{enumerate}
    \item First, we work out the \emph{gauge-fixing conditions} for the pre-geometric gauge potential that generate the different affine connections of metric-affine theories of gravity in the spontaneously broken phase.
    \item Secondly, we identify suitable pre-geometric \emph{actions} for the unbroken phase that allow for the emergence of more general metric-affine theories after the SSB.
\end{enumerate}
These two results conclusively prove that the metric-affine framework is fully compatible with the pre-geometric one, with the latter providing a unifying physical source for all the geometric notions of curvature, torsion and non-metricity. Our analysis focuses on the pre-geometric predecessors of the Geometric Trinity of Gravity first, and is then generalised to extensions thereof. Extra care is needed for the treatment of non-metricity because it requires dealing with a non-Lorentzian spin connection. From the point of view of gauge field theories, such as the pre-geometric ones, this difference is significant and necessitates an extension of the fundamental gauge group of spacetime; for this reason, we argue that this should be the five-dimensional general linear group whenever non-metricity is present in the emergent gravitational theory, while it is restricted to the (anti-)de Sitter group when the emergent linear connection is metric-compatible.

The notation used in this paper is the same as that of Ref.\ \cite{meluccio:pregeometric}. In short, internal indices are expressed as $A,B,C$ etc.\ in the unbroken phase and as $a,b,c$ etc.\ in the spontaneously broken phase, running from $0$ to $4$ and from $0$ to $3$ respectively; spacetime indices, instead, are denoted by $\lambda,\mu,\nu$ etc.\ and run from $0$ to $3$. The internal space metric $\eta$ that defines the gauge group $SO(1,4)$ or $SO(2,3)$ has signature $(-,+,+,+,\pm)$ respectively; in both cases, after a SSB that reduces the gauge group to $SO(1,3)$, $\eta$ becomes the Minkowski metric with signature $(-,+,+,+)$. Natural units are used throughout the paper, which is structured as follows. 

Sec.\ \ref{sec:2} provides a concise summary of both the metric-affine and the pre-geometric settings. Drawing upon the pre-geometric dictionary detailed in Sec.\ \ref{sec:3}, the relation between the gauge-fixing conditions for the pre-geometric field and the gauge choices for the metric-affine connection is then clarified in Sec.\ \ref{sec:4}. The pre-geometric versions of the actions of the Geometric Trinity of Gravity are presented in Sec.\ \ref{sec:5}, and further generalised in Sec.\ \ref{sec:6} to more general metric-affine theories. We conclude with some final remarks in Sec.\ \ref{sec:7}.

\section{Geometric and pre-geometric theories of gravity}\label{sec:2}
Metric-affine theories of gravity employ the Palatini formalism, thereby describing the gravitational field with two sets of independent variables: the components of the spacetime metric $g_{\mu\nu}$, on one hand, and those of the affine connection $\Gamma^\lambda_{\mu\nu}$, on the other hand. The metric can equivalently be replaced by the tetrad fields $e_\mu^a$ via the relation
\begin{equation}\label{eq:soldered}
    g_{\mu\nu}=\eta_{ab}e_\mu^ae_\nu^b.
\end{equation}
The affine connection $\Gamma^\lambda_{\mu\nu}$ and the spin connection $\omega_\mu^{ab}$ of the Lorentz group $SO(1,3)$ are related as
\begin{equation}\label{eq:spin-connection}
    \omega_\mu^{ab}=e_\lambda^a\partial_\mu e^{b\lambda}+e_\lambda^a\Gamma_{\mu\nu}^\lambda e^{b\nu}.
\end{equation}
The most general affine connection can in turn be decomposed as
\begin{equation}
    \Gamma_{\mu\nu}^\lambda=\mathring\Gamma_{\mu\nu}^\lambda+K_{\mu\nu}^{\phantom{\mu\nu}\lambda}+L_{\mu\nu}^{\phantom{\mu\nu}\lambda},
\end{equation}
with $\mathring\Gamma_{\mu\nu}^\lambda$ being the Levi-Civita connection; the contortion and disformation tensors are defined respectively as
\begin{align}
    K_{\mu\nu}^{\phantom{\mu\nu}\lambda}&\equiv\frac{1}{2}(T^\lambda_{\phantom{\lambda}\mu\nu}+T_{\nu\phantom{\lambda}\mu}^{\phantom{\nu}\lambda}-T_{\mu\nu}^{\phantom{\mu\nu}\lambda}),\\
    L_{\mu\nu}^{\phantom{\mu\nu}\lambda}&\equiv\frac{1}{2}(Q^\lambda_{\phantom{\lambda}\mu\nu}-Q_{\nu\phantom{\lambda}\mu}^{\phantom{\nu}\lambda}-Q_{\mu\nu}^{\phantom{\mu\nu}\lambda}),
\end{align}
where the torsion and non-metricity tensors are given respectively by
\begin{align}
    T_{\phantom{\lambda}\mu\nu}^\lambda&=2\Gamma^\lambda_{[\mu\nu]},\\
    Q_{\lambda\mu\nu}&=\partial_\lambda g_{\mu\nu}-2\Gamma_{\lambda(\mu}^\rho g_{\nu)\rho}.\label{eq:Q1}
\end{align}

Pre-geometric theories of gravity, instead, describe spacetime as a four-dimensional differentiable manifold lacking any metric structure. Rather, the spacetime metric emerges only as an effective field after a mechanism of SSB. This process is realised when a real, Higgs-like field $\phi^A$ acquires a nonzero vacuum expectation value $v$ along an internal space direction, say $\langle\phi^A\rangle=v\delta^A_4$. This construction implicitly assumes the existence of a pre-quantum Fock space, in order for the pre-geometric Higgs field to have a well-defined vacuum expectation value.  As a consequence of the SSB, the gauge symmetry of the theory is reduced, typically from the de Sitter group $SO(1,4)$ or the anti-de Sitter group $SO(2,3)$ to the Lorentz group $SO(1,3)$ (see Ref.\ \cite{addazi:pre-geometry} for more details). The components of the gauge field $A_\mu^{AB}$ of the (anti-)de Sitter group can then be identified as follows, depending on whether they contain the broken internal space direction or not:
\begin{equation}\label{eq:identifications}
    \omega_\mu^{ab}\xleftarrow{SSB}A_\mu^{ab},\qquad e_\mu^a\xleftarrow{SSB}m^{-1}A_\mu^{a4},
\end{equation}
where $m$ is a mass parameter introduced to render the tetrads dimensionless. The emergent spin connection represents the gauge connection of the residual symmetry group, i.e.\ the Lorentz group; the emergent tetrads correspond instead to the broken generators of the SSB. An exemplifying pre-geometric Lagrangian density is the one proposed by Wilczek \cite{wilczek:gauge}:\footnote{Another suitable Lagrangian density for the pre-geometry of spacetime is the one introduced by MacDowell and Mansouri \cite{macdowell:unified}. For studies of the MacDowell--Mansouri theory as a pre-geometric theory of gravity, the interested reader is referred to the works \cite{addazi:pre-geometry,addazi:hamiltonian,meluccio:pregeometric,addazi:emergence}.
Specifically, as argued in Ref.\cite{addazi:pre-geometry}, the Lagrangian density \eqref{eq:wilczek} is one of only two polynomial combinations of the pre-geometric fields and their derivatives which enables the recovery of the Hilbert–Einstein action after the mechanism of SSB. Its relevance is thus justified a posteriori via the correspondence principle. This is not merely a matter of simplicity, given that the high level of rigidity of the pre-geometric formalism (due to the absence of a spacetime metric) implies that the MacDowell–Mansouri action is the only other viable option; in other words, the configuration space of Pre-geometric Theories of Gravity is highly constrained.}

\begin{equation}\label{eq:wilczek}
    \mathcal{L}_\textup{W}\equiv k_\textup{W}\epsilon_{ABCDE}\epsilon^{\mu\nu\rho\sigma}F_{\mu\nu}^{AB}\nabla_\rho\phi^C\nabla_\sigma\phi^D\phi^E,
\end{equation}
where $k_\textup{W}$ is a nonzero constant with $[k_\textup{W}]=[\phi]^{-3}$. The field strength $F_{\mu\nu}^{AB}$ of the gauge field $A_\mu^{AB}$ and the related covariant derivative $\nabla_\mu$ are defined respectively as
\begin{align}
    F_{\mu\nu}^{AB}&=2\partial_{[\mu}A_{\nu]}^{AB}+2A_{\phantom{A}C[\mu}^AA_{\nu]}^{CB},\\
    \nabla_\mu\phi^A&=\partial_\mu\phi^A+A_{\phantom{A}B\mu}^A\phi^B.
\end{align}
When the fluctuations of the Higgs-like field around its vacuum expectation value are frozen out at low enough energies, the SSB of the Wilczek theory yields exactly the Einstein--Hilbert Lagrangian density with the cosmological constant term:
\begin{equation}\label{eq:L_W-SSB}
    \mathcal{L}_\textup{W}\xrightarrow{SSB}\mathcal{L}_\textup{EH}+\mathcal{L}_\Lambda\equiv\frac{M_\textup{P}^2}{2}eR-M_\textup{P}^2\Lambda e,
\end{equation}
where $e=\sqrt{-g}$ is the tetrad determinant and $R$ is the Ricci scalar. The emergent energy scales of GR, i.e.\ the (reduced) Planck mass and the cosmological constant, are identified respectively as
\begin{equation}\label{eq:emergent-energies}
     M_\textup{P}^2\equiv-8k_\textup{W}v^3m^2,\qquad\Lambda\equiv\pm6m^2,
\end{equation}
with the sign of the cosmological constant depending on whether the gauge group of the theory is $SO(1,4)$ or $SO(2,3)$ respectively.

Therefore, the specific pre-geometric Lagrangian density $\mathcal{L}_\textup{W}$ yields, in the spontaneously broken phase, a metric-affine theory of gravity based on the curvature invariant $R$ and with a generic metric-compatible connection; in other words, this is the Einstein--Cartan theory (see Ref.\ \cite{meluccio:pregeometric} for a more in-depth analysis). The aim of this paper is to generalise this result to pre-geometric theories whose SSB gives rise to more general metric-affine theories of gravity, including those based on torsion and non-metricity invariants, as well as to more general affine connections. Throughout this work, the fluctuations of the Higgs-like field around its vacuum expectation value after the SSB will be ignored; this means assuming that $\phi^A\xrightarrow{SSB}v\delta^A_4$. These fluctuations, in fact, quantify the deviations from the predictions of the emergent gravitational theory, which come in the form of the dynamics of an additional scalar field \cite{addazi:pre-geometry,addazi:hamiltonian,meluccio:pregeometric}; only once the latter is frozen out, the metric-affine theory is recovered exactly, and the signatures of its pre-geometric origins are effectively erased.

\section{Pre-geometric dictionary}\label{sec:3}
As reported in Ref.\ \cite{addazi:pre-geometry}, the following is a convenient way to express a pre-geometric dictionary for converting between pre-geometric quantities in the unbroken phase and emergent geometric quantities in the spontaneously broken phase:
\begin{equation}\label{eq:dictionary}
    \begin{split}
    \pm v^{-1}m^{-1}\nabla_\mu\phi^A&\xrightarrow{SSB}e_\mu^a,\\
    4vmJ^{-1}w_A^\mu&\xrightarrow{SSB}e_a^\mu,\\
    -\frac{1}{24}v^{-5}m^{-4}J&\xrightarrow{SSB}e=\sqrt{-g},\\
    v^{-2}m^{-2}P_{\mu\nu}&\xrightarrow{SSB}\eta_{ab}e_\mu^ae_\nu^b\equiv g_{\mu\nu},\\
    16v^2m^2M^{\mu\nu}&\xrightarrow{SSB}\eta^{ab}e_a^\mu e_b^\nu\equiv g^{\mu\nu},
    \end{split}
\end{equation}
where
\begin{align}
    P_{\mu\nu}&\equiv\eta_{AB}\nabla_\mu\phi^A\nabla_\nu\phi^B\xrightarrow{SSB}v^2m^2g_{\mu\nu},\\
    w_A^\mu&\equiv\pm\epsilon_{ABCDE}\epsilon^{\mu\nu\rho\sigma}\nabla_\nu\phi^B\nabla_\rho\phi^C\nabla_\sigma\phi^D\phi^E\nonumber\\
    &\xrightarrow{SSB}-6v^4m^3ee_a^\mu,\\
    J&\equiv\epsilon_{ABCDE}\epsilon^{\mu\nu\rho\sigma}\nabla_\mu\phi^A\nabla_\nu\phi^B\nabla_\rho\phi^C\nabla_\sigma\phi^D\phi^E\nonumber\\
    &\xrightarrow{SSB}-24v^5m^4e,\label{eq:J}\\
    M^{\mu\nu}&\equiv J^{-2}\eta^{AB}w_A^\mu w_B^\nu\xrightarrow{SSB}\frac{1}{16}v^{-2}m^{-2}\eta^{ab}e_a^\mu e_b^\nu.
\end{align}
The definition of the pre-geometric dictionary \eqref{eq:dictionary} requires using the Higgs-like field in order to define the desired pre-geometric quantities in a way that is (anti-)de Sitter-covariant. As a matter of fact, whilst it is possible to go without any $\phi^A$ or $\nabla_\mu\phi^A$ factors, the appearance of factors $A_\mu^{A4}$ in place of $\pm v^{-1}\nabla_\mu\phi^A$ or $\epsilon_{ABCD4}$ in place of $v^{-1}\epsilon_{ABCDE}\phi^E$ would then explicitly break such symmetry. This clarification is not unmotivated here. In the next section, in fact, we will be concerned with constructing suitable gauge-fixing conditions for the pre-geometric field $A_\mu^{AB}$, which need not be (anti-)de Sitter-covariant. For this purpose, we will adopt an alternative way to express the pre-geometric dictionary \eqref{eq:dictionary}, which is not (anti-)de Sitter-covariant but is built \emph{solely} out of the components of $A_\mu^{AB}$:
\begin{equation}
    \begin{split}
    m^{-1}A_\mu^{A4}&\xrightarrow{SSB}e_\mu^a,\\
    4m\tilde J^{-1}\tilde w_A^\mu&\xrightarrow{SSB}e_a^\mu,\\
    -\frac{1}{24}m^{-4}\tilde J&\xrightarrow{SSB}e=\sqrt{-g},\\
    m^{-2}\eta_{AB}A_\mu^{A4}A_\nu^{B4}&\xrightarrow{SSB}\eta_{ab}e_\mu^ae_\nu^b\equiv g_{\mu\nu},\\
    16m^2\tilde M^{\mu\nu}&\xrightarrow{SSB}\eta^{ab}e_a^\mu e_b^\nu\equiv g^{\mu\nu},
    \end{split}
\end{equation}
where
\begin{align}
    \tilde w_A^\mu&\equiv\epsilon_{ABCD4}\epsilon^{\mu\nu\rho\sigma}A_\nu^{B4}A_\rho^{C4}A_\sigma^{D4}\xrightarrow{SSB}-6m^3ee_a^\mu,\\
    \tilde J&\equiv\epsilon_{ABCD4}\epsilon^{\mu\nu\rho\sigma}A_\mu^{A4}A_\nu^{B4}A_\rho^{C4}A_\sigma^{D4}\xrightarrow{SSB}-24m^4e,\\
    \tilde M^{\mu\nu}&\equiv\tilde J^{-2}\eta^{AB}\tilde w_A^\mu\tilde w_B^\nu\xrightarrow{SSB}\frac{1}{16}m^{-2}\eta^{ab}e_a^\mu e_b^\nu.
\end{align}
Henceforth we will use the shorthand notation $A^{AB\mu}\equiv16m^2\tilde M^{\mu\nu}A_\nu^{AB}$, whose explicit expression is
\begin{equation}
    \begin{split}
        A^{AB\mu}&=16m^2\eta^{CD}\epsilon_{CIJK4}\epsilon_{DLMN4}\epsilon^{\mu\alpha\beta\gamma}\epsilon^{\nu\delta\xi\zeta}\\
        &\cdot(\epsilon_{EFGH4}\epsilon^{\lambda\rho\sigma\tau}A_\lambda^{E4}A_\rho^{F4}A_\sigma^{G4}A_\tau^{H4})^{-2}\\
        &\cdot A_\alpha^{I4}A_\beta^{J4}A_\gamma^{K4}A_\delta^{L4}A_\xi^{M4}A_\zeta^{N4}A_\nu^{AB}.
    \end{split}
\end{equation}
This definition implies that $A^{ab\mu}\xrightarrow{SSB}g^{\mu\nu}\omega_\nu^{ab}=\omega^{ab\mu}$.

\section{Affine geometry from gauge-fixing}\label{sec:4}
In this section we set out to derive the pre-geometric version of the most common gauge conditions used in metric-affine theories of gravity. In particular, we will reconstruct the Levi-Civita, the Weitzenb\"{o}ck and the coincident gauges from specific combinations of the components of the pre-geometric field $A_\mu^{AB}$ and their first derivatives.

The Riemann curvature tensor, which is defined as
\begin{equation}
    R_{\mu\nu}^{ab}=2\partial_{[\mu}\omega_{\nu]}^{ab}+2\omega_{\phantom{a}c[\mu}^a\omega_{\nu]}^{cb},
\end{equation}
can be obtained in the pre-geometric framework from the SSB of the field strength of $A_\mu^{AB}$:
\begin{equation}\label{eq:field-strength}
    F_{\mu\nu}^{AB}\xrightarrow{SSB}R_{\mu\nu}^{ab}\mp2m^2e_{[\mu}^ae_{\nu]}^b.
\end{equation}
The Levi-Civita connection of GR is the unique torsionless and metric-compatible connection on a pseudo-Riemannian manifold, and it is completely specified in terms of the metric and its first derivatives:
\begin{equation}
    \mathring\Gamma_{\mu\nu}^\lambda=\frac{1}{2}g^{\lambda\rho}(\partial_\mu g_{\rho\nu}+\partial_\nu g_{\mu\rho}-\partial_\rho g_{\mu\nu}).
\end{equation}
Making use of Eqs.\ \eqref{eq:soldered} and \eqref{eq:spin-connection}, the related spin connection can be expressed entirely in terms of the tetrads and their first derivatives:
\begin{equation}
    \begin{split}
        \mathring\omega_\mu^{ab}&=e_\lambda^a\partial_\mu e^{b\lambda}+e_\lambda^a\mathring\Gamma_{\mu\nu}^\lambda e^{b\nu}\\
        &=e^{[a|\nu|}(\partial_\mu e_\nu^{b]}-\partial_\nu e_\mu^{b]}+e^{b]\lambda}e_{c\mu}\partial_\lambda e_\nu^c).
    \end{split}
\end{equation}
The corresponding gauge-fixing condition in the pre-geometric formalism is then
\begin{equation}
    \begin{split}
        A_\mu^{ab}&=m^{-2}A^{[a|4\nu|}(\partial_\mu A_\nu^{b]4}-\partial_\nu A_\mu^{b]4}-m^{-2}A^{b]4\lambda}A_{\phantom{4}c\mu}^4\partial_\lambda A_\nu^{c4})\\
        &\xrightarrow{SSB}\omega_\mu^{ab}=\mathring\omega_\mu^{ab}.
\end{split}
\end{equation}

Next we consider the torsion tensor, which in terms of the spin connection and the tetrad fields is expressed (using Eqs.\ \eqref{eq:soldered} and \eqref{eq:spin-connection}) as
\begin{equation}
    T_{\mu\nu}^a=2\mathcal{D}_{[\mu}e_{\nu]}^a,
\end{equation}
where $\mathcal{D}_\mu$ is the covariant derivative with respect to the spin connection. In pre-geometric terms, this tensor can be reconstructed after the SSB as follows:
\begin{equation}\label{eq:torsion}
    \pm2v^{-1}m^{-1}\nabla_{[\mu}\nabla_{\nu]}\phi^A\xrightarrow{SSB}T_{\mu\nu}^a.
\end{equation}
The Weitzenb\"{o}ck connection of TEGR is a flat and metric-compatible connection given by
\begin{equation}
    \hat\Gamma_{\mu\nu}^\lambda=e_a^\lambda\partial_\mu e_\nu^a.
\end{equation}
It results from the Weitzenb\"{o}ck gauge condition, i.e.\ the vanishing of the spin connection:
\begin{equation}
    \hat\omega_\mu^{ab}=0.
\end{equation}
The pre-geometric formulation of the Weitzenb\"{o}ck gauge is thus simply
\begin{equation}
    A_\mu^{ab}=0\xrightarrow{SSB}\omega_\mu^{ab}=\hat\omega_\mu^{ab}.
\end{equation}

Finally, the definition of the non-metricity tensor can be recast (again, using Eqs.\ \eqref{eq:soldered} and \eqref{eq:spin-connection}) as
\begin{equation}\label{eq:Q2}
    Q_\mu^{ab}=\mathcal{D}_\mu\eta^{ab}=2\omega_\mu^{(ab)}.
\end{equation}
This expression highlights the fact that non-metricity, unlike curvature or torsion, is not defined via the spin connection of the Lorentz group but rather via that of the general linear group. This is because while the spin connection of $SO(1,3)$ is antisymmetric in its internal indices, that of $GL(4)$ is not and thus admits also a symmetric part. This important difference reflects on the construction for the unbroken phase too. Indeed, the non-metricity tensor can be recovered from the SSB of the following pre-geometric quantity:
\begin{equation}\label{eq:non-metricity}
    \nabla_\mu\eta^{AB}=2A_\mu^{(AB)}\xrightarrow{SSB}Q_\mu^{ab},\qquad A,B\ne4.
\end{equation}
Again, this expression shows that $A_\mu^{AB}$ cannot be the gauge potential of the (anti-)de Sitter group, for that is antisymmetric in its internal indices; rather, it must be generalised to the gauge potential of a broader group, like the general linear group $GL(5)$. In Eq.\ \eqref{eq:non-metricity}, $\eta$ is the constant symbol $\eta^{AB}=\diag(-1,1,1,1,1)$ whose gauge covariant derivative quantifies the deviations from the metric-compatibility condition at any spacetime point. Hence, $\eta$ does not represent a fixed internal space background, both because that is not defined for the $GL(5)$ group (differently from the case of $SO(1,4)$ or $SO(2,3)$) and because it would be at odds with the spirit of pre-geometric gravity; the latter, in fact, completely relinquishes the notion of a fundamental dynamical metric, both for the spacetime manifold and for its tangent bundle. When it comes to finding a convenient gauge, in this case the coincident gauge of STEGR yields a spin connection that is completely specified by the tetrads, specifically
\begin{equation}
    \stg\omega_\mu^{ab}=-e^{b\lambda}\partial_\mu e_\lambda^a,
\end{equation}
and is due to the vanishing of the affine connection:
\begin{equation}
    \stg\Gamma_{\mu\nu}^\lambda=0.
\end{equation}
The pre-geometric version of the coincident gauge is
\begin{equation}
    A_\mu^{ab}=-m^{-2}A^{b4\lambda}\partial_\mu A_\lambda^{a4}\xrightarrow{SSB}\omega_\mu^{ab}=\stg\omega_\mu^{ab}.
\end{equation}
The expressions of the torsion and non-metricity tensors simplify considerably when adopting, respectively, the Weitzenb\"{o}ck and coincident gauges:
\begin{equation}
    \hat T_{\mu\nu}^a=2\partial_{[\mu}e_{\nu]}^a,\qquad \stg Q_{\lambda\mu\nu}=\partial_\lambda g_{\mu\nu}.
\end{equation}
We conclude this section by remarking that the Higgs-like field $\phi^A$ is responsible for reducing the fundamental gauge symmetry of spacetime from $SO(1,4)$ or $SO(2,3)$ to $SO(1,3)$ in pre-geometric theories corresponding to gravitational theories based on curvature or torsion; instead, the same mechanism of SSB must be understood as yielding the gauge group $GL(4)$ from a broader group, like $GL(5)$, in the case of pre-geometric theories corresponding to gravitational theories based on non-metricity. This subtle difference will be expanded on in the following section.

\section{Gravitational dynamics from pre-geometry}\label{sec:5}
The actions of Geometric Trinity of Gravity are based on three different geometric invariants, namely the Ricci scalar $R$, the torsion scalar $T$ and the non-metricity scalar $Q$. Moreover, each of these utilises a specific gauge condition for the affine connection. We will now focus on constructing the pre-geometric version of these gravitational actions, with the correspondence between geometric and pre-geometric formalisms arising as a consequence of the SSB mechanism elucidated in Sec.\ \ref{sec:2}. As for the implementation of any gauge condition for a given pre-geometric theory, it is understood that this can be achieved via the method outlined in the previous section, so as to recover the desired type of affine connection in the spontaneously broken phase. Whenever a Lagrangian density for the unbroken phase is presented next, the corresponding action is given simply by its integral over the coordinates of the whole four-dimensional spacetime manifold.

In Sec.\ \ref{sec:2}, it was shown how to recover GR's Einstein--Hilbert action and the cosmological constant term -- with a generic metric-compatible connection -- starting from the pre-geometric setting (see Eq.\ \eqref{eq:L_W-SSB}). For completeness, here we will build a pre-geometric action for obtaining the same result but without the cosmological constant term. The most straightforward way to do this is to add a term proportional to $J$ (defined in Eq.\ \eqref{eq:J}) to the Wilczek Lagrangian density (Eq.\ \eqref{eq:wilczek}):
\begin{equation}\label{eq:L_R}
    \mathcal{L}_R\equiv\mathcal{L}_\textup{W}+k_JJ,
\end{equation}
with $[k_J]=[\phi]^{-5}$. If $k_J=\pm2k_\textup{W}/v^2$, in fact, the SSB of the term $k_JJ$ will exactly cancel the contribution $\mathcal{L}_\Lambda$ coming from the SSB of $\mathcal{L}_\textup{W}$, thereby yielding
\begin{equation}\label{eq:EH}
    \mathcal{L}_R\xrightarrow{SSB}\mathcal{L}_\textup{EH}\equiv\frac{M_\textup{P}^2}{2}eR,
\end{equation}
with the emergent Planck mass given by Eq.\ \eqref{eq:emergent-energies}. A more elaborate but equally valid pre-geometric version for the Einstein--Hilbert Lagrangian density can be found by replacing $\mathcal{L}_\textup{W}$ in Eq.\ \eqref{eq:L_R} with an alternative expression as follows:
\begin{equation}
    \mathcal{L}_R\equiv k_RJ^{-1}w^\mu_Aw^\nu_BF_{\mu\nu}^{AB}+k_JJ,
\end{equation}
where $[k_R]=[\phi]^{-3}$. This time, setting $k_J=\pm3k_R/4v^2$ and using Eqs.\ \eqref{eq:dictionary} and \eqref{eq:field-strength} leads to the same result \eqref{eq:EH} but with an emergent Planck mass defined as
\begin{equation}
    M_\textup{P}^2\equiv-3k_Rv^3m^2.
\end{equation}

We now turn to the problem of constructing the pre-geometric form of the TEGR action, which is based on the torsion scalar. This invariant is given by
\begin{equation}
    T\equiv\frac{1}{4}T^\lambda_{\phantom{\lambda}\mu\nu}T_\lambda^{\phantom{\lambda}\mu\nu}+\frac{1}{2}T^\lambda_{\phantom{\lambda}\mu\nu}T^{\nu\mu}_{\phantom{\nu\mu}\lambda}-T^\mu T_\mu,
\end{equation}
where $T^\mu\equiv T^{\lambda\mu}_{\phantom{\lambda\mu}\lambda}$ is the torsion vector. In the unbroken phase we can define the following scalars:
\begin{equation}\label{eq:T_1,2,3}
    \begin{split}
        T_1&\equiv1024v^2m^2\eta_{AB}M^{\mu\rho}M^{\nu\sigma}\nabla_{[\mu}\nabla_{\nu]}\phi^A\nabla_{[\rho}\nabla_{\sigma]}\phi^B,\\
        T_2&\equiv1024v^2m^2J^{-2}w^\sigma_Aw^\nu_BM^{\mu\rho}\nabla_{[\mu}\nabla_{\nu]}\phi^A\nabla_{[\rho}\nabla_{\sigma]}\phi^B,\\
        T_3&\equiv1024v^2m^2J^{-2}w^\sigma_Aw^\nu_BM^{\mu\rho}\nabla_{[\mu}\nabla_{\nu]}\phi^B\nabla_{[\rho}\nabla_{\sigma]}\phi^A.
    \end{split}
\end{equation}
Using Eqs.\ \eqref{eq:dictionary} and \eqref{eq:torsion}, it is easy to prove that
\begin{equation}
    \begin{split}
        T_1\xrightarrow{SSB}\eta_{ab}g^{\mu\rho}g^{\nu\sigma}T^a_{\phantom{a}\mu\nu}T^b_{\phantom{b}\rho\sigma}=T^a_{\phantom{a}\mu\nu}T_a^{\phantom{a}\mu\nu}=T^\lambda_{\phantom{\lambda}\mu\nu}T_\lambda^{\phantom{\lambda}\mu\nu},\\
        T_2\xrightarrow{SSB}e^\sigma_ae^\nu_bg^{\mu\rho}T^a_{\phantom{a}\mu\nu}T^b_{\phantom{b}\rho\sigma}=T^a_{\phantom{a}\mu\nu}T^{\nu\mu}_{\phantom{\nu\mu}a}=T^\lambda_{\phantom{\lambda}\mu\nu}T^{\nu\mu}_{\phantom{\nu\mu}\lambda},\\
        T_3\xrightarrow{SSB}e^\sigma_ae^\nu_bg^{\mu\rho}T^b_{\phantom{b}\mu\nu}T^a_{\phantom{a}\rho\sigma}=T^{a\mu}_{\phantom{a\mu}a}T^b_{\phantom{b}\mu b}=T^\mu T_\mu.
    \end{split}
\end{equation}
Hence, it follows that
\begin{equation}
    \frac{1}{4}T_1+\frac{1}{2}T_2-T_3\xrightarrow{SSB}T.
\end{equation}
Therefore, the pre-geometric version of the TEGR Lagrangian density can be defined as
\begin{equation}\label{eq:L_T}
    \mathcal{L}_T\equiv k_TJ\biggl(\frac{1}{4}T_1+\frac{1}{2}T_2-T_3\biggr),
\end{equation}
with $[k_T]=[m]^{-2}[\phi]^{-5}$. To recover the TEGR Lagrangian density after the SSB, that is
\begin{equation}
    \mathcal{L}_T\xrightarrow{SSB}\mathcal{L}_\textup{TEGR}\equiv\frac{M_\textup{P}^2}{2}eT,
\end{equation}
the emergent Planck mass must be identified as
\begin{equation}
    M_\textup{P}^2\equiv-48k_Tv^5m^4.
\end{equation}
Note that the Lagrangian density \eqref{eq:L_T} does not contain second-order time derivatives of the scalar field $\phi^A$ -- thanks to the antisymmetrisation of spacetime indices in the expressions \eqref{eq:T_1,2,3} -- and is thus consistent with the absence of Ostrogradsky instabilities \cite{woodard:ostrogradsky}.

STEGR is based on the non-metricity scalar, namely
\begin{equation}
    Q\equiv g^{\mu\nu}(L_{\sigma\mu}^{\phantom{\sigma\mu}\rho}L_{\nu\rho}^{\phantom{\nu\rho}\sigma}-L_{\sigma\rho}^{\phantom{\sigma\rho}\rho}L_{\mu\nu}^{\phantom{\mu\nu}\sigma}).
\end{equation}
The steps needed for constructing the pre-geometric form of the STEGR action are similar to those already seen in this section, but the precise model building hinges on the specifics of the implementation of the SSB mechanism for bridging the unbroken and the spontaneously broken phases. As anticipated in Sec.\ \ref{sec:4}, indeed, the introduction of non-metricity requires the generalisation of the gauge group of any metric-affine theory from the Lorentz group to the general linear group. This feature, in fact, enables the possibility of considering a spin connection that is not antisymmetric in its internal indices, which in turn leads to a non-trivial metric-compatibility constraint for the affine connection (see Eqs.\ \eqref{eq:Q1} and \eqref{eq:Q2}). Before proceeding further, a reflection on results from group theory is thus necessary. For pre-geometric theories that recover a metric-compatible connection ($Q=0$) after the SSB, the fundamental gauge group of spacetime can be either $SO(1,4)$ or $SO(2,3)$, whose $10$ generators decompose exactly into the $6$ generators of $SO(1,3)$ and the $4$ generators of the translation group $\mathbb{R}^4$. As for pre-geometric theories that recover non-metricity ($Q\ne0$) after the SSB, instead, in order to obtain $GL(4)$ in the spontaneously broken phase a natural expectation for the fundamental gauge group of spacetime would be $GL(5)$. However, the $25$ generators of $GL(5)$ do not decompose only into the $16$ generators of $GL(4)$ and the $4$ generators of the translation group $\mathbb{R}^4$, but also into the $4$ generators of the dual representation $\mathbb{R}^{*4}$ of the translation group and the single generator of dilations. More generally, it can be proven (see Ref.\ \cite{hehl:metric} and references therein) that the semidirect product of four-dimensional translations and general linear transformations cannot be obtained from a group contraction of the five-dimensional group of general linear transformations, nor any other semisimple group. Anyway, in gauge field theories such as the pre-geometric ones, the gauge freedom for the gauge potential can be exploited for cancelling undesired degrees of freedom. In the case of pre-geometric theories with $GL(5)\xrightarrow{SSB}GL(4)$, in particular, the gauge-fixing condition that ensures the exact recovery of the gravitational degrees of freedom in the presence of non-metricity is
\begin{equation}\label{eq:gauge-Q}
    \bar{A}_\mu^{4A}=0.
\end{equation}
In doing so, indeed, the remaining nonzero components of the gauge-fixed potential $\bar{A}_\mu^{AB}$ split into the $16$ components $\bar{A}_\mu^{ab}$ of the non-Lorentzian spin connection of $GL(4)$ and the $4$ components $\bar{A}_\mu^{a4}$ corresponding to the tetrads, via the same identifications shown in Eq.\ \eqref{eq:identifications}. From a mathematical point of view, the above condition, though simple and conceptually clear, is not unique, given that the same outcome after the SSB may still be obtained with different relations through adequate redefinitions of the emerging geometric fields (in place of the definitions \eqref{eq:identifications}). Henceforth a bar over any pre-geometric quantity will signify that all the gauge potentials it implicitly contains are gauge-fixed with the condition \eqref{eq:gauge-Q}. This gauge-fixing condition must be assumed as a requisite for the construction of pre-geometric theories that lead to gravitational theories sourced by non-metricity; conceptually this is in contrast to all the gauge-fixing conditions discussed in Sec.\ \ref{sec:4}, which are, in a sense, accessory and useful only insofar as they simplify the formalism. In any case, it is certainly possible to envision alternative mechanisms for a SSB that yields the $GL(4)$ group and the tetrads in the context of a pre-geometric theory of gravity, and the consequences of this important choice in model building deserve further investigations. We are now ready to resume the construction of a pre-geometric theory that gives rise to the non-metricity scalar after the SSB. To do so, in the unbroken phase one can define the tensor
\begin{equation}
    \begin{split}
        \bar{N}_{\mu\nu}^{\phantom{\mu\nu}\lambda}\equiv8&\eta_{AC}\eta_{BD}\bar{M}^{\lambda\sigma}\bar{\nabla}_\mu\phi^C\bar{\nabla}_\nu\phi^D\bar{\nabla}_\sigma\eta^{AB}\\
        &\mp4\eta_{BC}\bar{J}^{-1}\bar{w}^\lambda_A\bar{\nabla}_{(\nu}\phi^C\bar{\nabla}_{\mu)}\eta^{AB},
    \end{split}
\end{equation}
whose SSB yields, via the Eqs.\ \eqref{eq:dictionary} and \eqref{eq:non-metricity}, exactly the disformation tensor:
\begin{equation}
    \bar{N}_{\mu\nu}^{\phantom{\mu\nu}\lambda}\xrightarrow{SSB}L_{\mu\nu}^{\phantom{\mu\nu}\lambda}.
\end{equation}
The pre-geometric version of the STEGR Lagrangian density is thus
\begin{equation}
    \mathcal{L}_Q\equiv k_Q\bar{J}\bar{M}^{\mu\nu}(\bar{N}_{\sigma\mu}^{\phantom{\sigma\mu}\rho}\bar{N}_{\nu\rho}^{\phantom{\nu\rho}\sigma}-\bar{N}_{\sigma\rho}^{\phantom{\sigma\rho}\rho}\bar{N}_{\mu\nu}^{\phantom{\mu\nu}\sigma}),
\end{equation}
with $[k_Q]=[\phi]^{-3}$. This time the STEGR Lagrangian density is recovered after the SSB, i.e.\
\begin{equation}
    \mathcal{L}_Q\xrightarrow{SSB}\mathcal{L}_\textup{STEGR}\equiv\frac{M_\textup{P}^2}{2}eQ,
\end{equation}
if the emergent Planck mass is identified as
\begin{equation}
    M_\textup{P}^2\equiv-3k_Qv^3m^2.
\end{equation}

\section{More general pre-geometric theories}\label{sec:6}
It is straightforward to generalise the constructions presented in the previous section to the pre-geometric analogue of various $f(R,T,Q)$ theories, where $f$ is a sufficiently smooth function (see also Refs.\ \cite{harko:non-minimal,iosifidis:metric-affine,iosifidis:vector-tensor}). To reproduce curvature-based theories, for example, a possible Lagrangian density for the unbroken phase is
\begin{equation}
    \mathcal{L}_{f(R)}\equiv Jf(J^{-1}\mathcal{L}_R)\xrightarrow{SSB}k_{f(R)}ef(R),
\end{equation}
with $k_{f(R)}$ being a nonzero constant. In the case $f(R)=R^n$, in particular, the SSB is given explicitly by
\begin{equation}
    \mathcal{L}_{R^n}=J(J^{-1}\mathcal{L}_\textup{W}+k_J)^n\xrightarrow{SSB}k_{R^n}eR^n,
\end{equation}
with $k_J=\pm2k_\textup{W}/v^2$ and $k_{R^n}=-24k_\textup{W}^nv^{5-2n}m^{4-2n}/6^n$. Note that this prescription is consistent with the definition \eqref{eq:emergent-energies} of the emergent constant $M_\textup{P}^2$ for the case $n=1$, which yields $\mathcal{L}_\textup{EH}$ in the spontaneously broken phase.

For torsion-based theories, a suitable Lagrangian density for the unbroken phase is
\begin{multline}
    \mathcal{L}_{f(T_1,T_2,T_3)}\equiv k_{f(T_1,T_2,T_3)}Jf(T_1,T_2,T_3)\\
    \xrightarrow{SSB}k'_{f(T_1,T_2,T_3)}ef(T^\lambda_{\phantom{\lambda}\mu\nu}T_\lambda^{\phantom{\lambda}\mu\nu},T^\lambda_{\phantom{\lambda}\mu\nu}T^{\nu\mu}_{\phantom{\nu\mu}\lambda},T^\mu T_\mu),
\end{multline}
with $k'_{f(T_1,T_2,T_3)}=-24k_{f(T_1,T_2,T_3)}v^5m^4$ and $k_{f(T_1,T_2,T_3)}$ a nonzero constant. For instance, the case $f(T^\lambda_{\phantom{\lambda}\mu\nu}T_\lambda^{\phantom{\lambda}\mu\nu},T^\lambda_{\phantom{\lambda}\mu\nu}T^{\nu\mu}_{\phantom{\nu\mu}\lambda},T^\mu T_\mu)=f(T)$ can be recovered by setting $f(T_1,T_2,T_3)=f(T_1/4+T_2/2-T_3)$.

Similarly, for theories based on non-metricity, a possible definition for the Lagrangian density before the SSB is
\begin{multline}
    \mathcal{L}_{f(Q)}\equiv k_{f(Q)}\bar{J}f(\bar{M}^{\mu\nu}(\bar{N}_{\sigma\mu}^{\phantom{\sigma\mu}\rho}\bar{N}_{\nu\rho}^{\phantom{\nu\rho}\sigma}-\bar{N}_{\sigma\rho}^{\phantom{\sigma\rho}\rho}\bar{N}_{\mu\nu}^{\phantom{\mu\nu}\sigma}))\\
    \xrightarrow{SSB}k'_{f(Q)}ef(Q),
\end{multline}
where $k_{f(Q)}$ and $k'_{f(Q)}$ are two nonzero constants linked by a relation which depends on $f$.

Finally, putting together all the results of this section allows to write down pre-geometric Lagrangian densities which can reproduce gravitational theories based on multiple types of geometric invariants after the SSB. We will limit ourselves to sketching a pre-geometric version of $f(R,T,Q)$ theories:
\begin{multline}
    \mathcal{L}_{f(R,T,Q)}\equiv k_{f(R,T,Q)}\bar{J}f(\bar{J}^{-1}\bar{\mathcal{L}}_R,\bar{T}_1/4\\
    +\bar{T}_2/2-\bar{T}_3,\bar{M}^{\mu\nu}(\bar{N}_{\sigma\mu}^{\phantom{\sigma\mu}\rho}\bar{N}_{\nu\rho}^{\phantom{\nu\rho}\sigma}-\bar{N}_{\sigma\rho}^{\phantom{\sigma\rho}\rho}\bar{N}_{\mu\nu}^{\phantom{\mu\nu}\sigma}))\\
    \xrightarrow{SSB}k'_{f(R,T,Q)}ef(R,T,Q),
\end{multline}
where, again, $k_{f(R,T,Q)}$ and $k'_{f(R,T,Q)}$ are two nonzero constants linked by a relation which depends on $f$.

\begin{figure*}[t]
    \centering
    \includegraphics[scale=0.45]{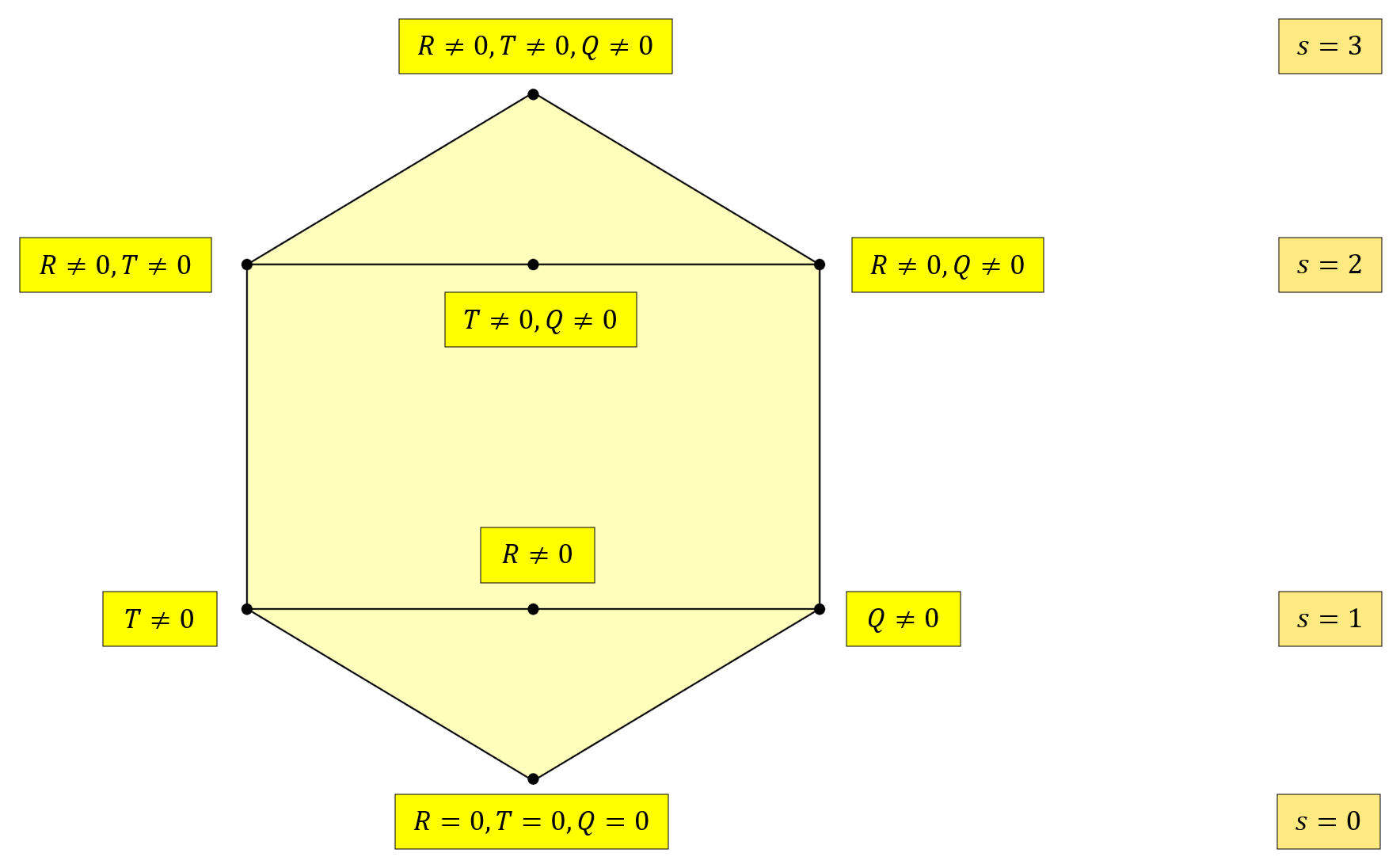}
    \caption{The octet of gravity, i.e.\ a diagrammatic classification of metric-affine theories of gravity based on the number $s$ of nonzero geometric invariants ($R$, $T$ or $Q$). The Geometric Trinity of Gravity corresponds to the set of theories with $s=1$. Figure adapted from Ref.\ \cite{capozziello:extended}.\label{fig:octet}}
\end{figure*}

\section{Conclusions and perspectives}\label{sec:7}
The aim of this paper is to establish a complete and consistent correspondence between metric-affine and pre-geometric theories of gravity. Any geometric theory of gravity, whose metric and affine structures give rise to the geometric invariants of curvature, torsion or non-metricity, can be recovered as an effective theory within the framework of pre-geometry. The emergence of the gravitational degrees of freedom is thus understood to stem from the SSB that reduces the fundamental gauge symmetry of pre-geometric spacetimes. This outcome sheds even more light on the deep relation between the geometric structure of spacetime, its pre-geometric origin and the degrees of freedom of the emergent gravitational field.

Our discussion centres around two key theoretical aspects: gauge choices and actions. The gauge choices for the affine connection of the metric-affine formalism are seen as arising from gauge-fixing conditions for the gauge potential of pre-geometric setting. The important difference lies in the fact that such conditions are not imposed on two independent set of variables, i.e.\ the metric (or the tetrads) and the affine connection, but rather on a single gauge potential, which is typically that of the (anti-)de Sitter group. This makes pre-geometric theories, in a sense, more natural than metric-affine ones, especially with reference to their formulation \textit{à la} Yang--Mills. Such a peculiar feature, far from being purely aesthetic, could have important implications for the development a fully-fledged theory of Quantum Gravity (see also Ref.\ \cite{addazi:hamiltonian}). As for the construction of suitable actions, the pre-geometric framework is found to be able of accommodating all the geometric invariants of metric-affine theories of gravity in its spontaneously broken phase. Out of the three possible geometric invariants of spacetime, only that of non-metricity requires a less straightforward elaboration in pre-geometric terms, because of the non-Lorentzian spin connection that it entails; as a consequence, the right gauge group for the unbroken phase cannot be the (anti-)de Sitter one in this case, nevertheless a natural generalisation is provided by the group of five-dimensional general linear transformations. Our analysis, as a matter of fact, is not even limited to the three theories of the Geometric Trinity of Gravity, because it applies equally well to any extensions thereof (like $f(R,T,Q)$ theories). Every pre-geometric action discussed in this article is indeed gauge invariant, generally covariant, background independent and, at least for the Geometric Trinity of Gravity, also free of Ostrogradsky instabilities. For completeness, we add that the method for introducing matter couplings in pre-geometric theories based on the (anti-)de Sitter group was presented in Ref.\ \cite{meluccio:pregeometric}, and it can be further generalised in the presence of non-metricity following the constructions elucidated in Ref.\ \cite{hehl:metric}.

One of the most interesting continuations of the present study would be carrying out the Hamiltonian analysis of all these pre-geometric theories, in order to address the properties of their algebra of constraints and the nature of their degrees of freedom. This is particularly true for those pre-geometric theories that involve non-metricity, for they rely on the implementation of a specific gauge-fixing condition (see, for example, Ref.\ \cite{yang:data}); if left unconstrained, instead, these theories will contain additional degrees of freedom other than those classically associated with the gravitational field. Also, an exhaustive list of the possible patterns of SSB for pre-geometric theories that generate non-metricity, though beyond the scope of the current article, will surely provide valuable insights into how flexible and all-encompassing the pre-geometric paradigm can be. Speaking of which, the pre-geometric theories that lead to the emergence of curvature or torsion with a metric-compatible affine connection definitely seem more natural, given the perfect role played by the (anti-)de Sitter group in generating the gravitational degrees of freedom after the SSB; that said, another contribution of our work is highlighting that pre-geometric gravity is a much wider research program than (anti-)de Sitter gravity, since the formalism of the former can include but is not limited to that of the latter.

Regarding phenomenological aspects, it should be kept in mind that, as discussed in Refs.\ \cite{addazi:pre-geometry,addazi:hamiltonian,meluccio:pregeometric,addazi:topological}, a pre-geometric theory of gravity like the Wilczek one possesses three degrees of freedom, whereas GR has only two (those of a massless graviton). The additional degree of freedom is represented by the scalar field $\phi^A$ and can be excited in a ultra-high-energy regime, which is expected to be close to the Planck scale. Just like this feature can lead to falsifiable deviations from the predictions of GR after the SSB, the same reasoning can ostensibly be extended to the other pre-geometric theories that produce the Geometric Trinity of Gravity in the spontaneously broken phase. Some of the most appealing predictions, in this sense, are related to the possibility of curing these theories' ultraviolet divergences, which are of geometric nature, through the dynamics of the additional degrees of freedom that are peculiar to pre-geometric spacetimes \cite{meluccio:pregeometric}.

We conclude with a reflection on the ``eightfold way of gravity'', a fascinating insight put forth in Ref.\ \cite{capozziello:extended}. In analogy with the particle multiplets of the Standard Model and the concept of quantum number, metric-affine theories of gravity can also be classified in terms of a number $s$ that counts the nonzero geometric invariants upon which each gravitational theory is based (curvature, torsion or non-metricity); equivalently, $s$ can be taken to be the number of non-trivial geometric properties of the affine connection (flatness, symmetry or metric-compatibility). The diagram for the classification of all metric-affine theories is therefore like an `octet of gravity' (see Fig.\ \ref{fig:octet}):
\begin{itemize}
    \item $s=0$ corresponds to Special Relativity, i.e.\ no gravity (as $R=T=Q=0$);
    \item $s=1$ comprises the three theories of Geometric Trinity of Gravity, namely GR (for $R\ne0$), TEGR (for $T\ne0$) and STEGR (for $Q\ne0$);
    \item $s=2$ includes, for example, the Einstein--Cartan theory ($R\ne0$ and $T\ne0$) \cite{hehl:spin} and the General Teleparallel Equivalent of General Relativity ($T\ne0$ and $Q\ne0$) \cite{aguiar-gomes:general,heisenberg:homogeneous,beltran-jimenez:general,bahamonde:teleparallel,bajardi:primary};
    \item $s=3$ refers to the case of the most general metric-affine theories ($R\ne0$, $T\ne0$ and $Q\ne0$).
\end{itemize}
In view of this configuration space, the results achieved in this work reveal that the five-dimensional general linear group could be the unifying symmetry group of all metric-affine theories -- when these are understood as emergent from the broader framework of pre-geometric gravity.

The pre-geometric approach considered in the present work is restricted to four dimensions. However higher dimensional formulations {\it à la} Kaluza-Klein, for example, are possible in this framework,
though a full discussion lies beyond the scope of this work. Ref.\cite{addazi:pre-geometry}, in particular, contains an introductory discussion about this  research direction. Some of the critical features to be studied for higher-dimensional pre-geometric theories are the choice of the gauge group and the construction of suitable actions (for example, extensions of the (anti-)de Sitter group would require employing Levi-Civita symbols with more than 5 internal indices). In a forthcoming paper, these topics  will be discussed in detail.

\acknowledgments
The authors thank Andrea Addazi and Daniel Blixt for comments and suggestions that allowed to improve the manuscript, and acknowledge the support of Istituto Nazionale di Fisica Nucleare (INFN), Sezione di Napoli, \textit{Iniziative Specifiche} QGSKY and MoonLIGHT-2. This publication is based upon work from COST Action CA21136 -- ``Addressing observational tensions in cosmology with systematics and fundamental physics (CosmoVerse)'', supported by COST (European Cooperation in Science and Technology).

\end{document}